# Flux Jumping and a Bulk-to-Granular Transition in the Magnetization of a Compacted and Sintered $MgB_2$ Superconductor

S.X. Dou, X.L. Wang, J. Horvat, D. Milliken, E.W. Collings* and M. D. Sumption*
Institute of Superconducting and Electronic Materials, University of Wollongong, Northfields Ave. Wollongong, NSW 2522 Australia

*LASM, Department of Materials Science and Engineering, The Ohio State University, College Ave, Columbus, OH 43210-1179, USA

## Abstract

In this letter, we report the results of field (H) and temperature (T) dependent magnetization (M) measurements of a pellet of uniform, large-grain sintered $MgB_2$. We show that at low temperatures the size of the pellet and its critical current density, $J_c$(H) -- i.e. its M(H) -- ensure low field flux jumping, which of course ceases when M(H) drops below a critical value. With further increase of H and T the individual grains decouple and the M(H) loops drop to lower lying branches, unresolved in the usual full M(H) representation. After taking into account the sample size and grain size, respectively, the bulk sample and the grains were deduced to exhibit the same magnetically determined $J_c$ s (e.g. $10^5$ A/cm$^2$, 20 K, 0T) and hence that for each temperature of measurement $J_c$(H) decreased monotonically with H over the entire field range, except for a gap within the grain-decoupling zone.

## I. Introduction

The recent discovery of intermediate-temperature superconductivity (ITC) in MgB2 by Akimitsu et al. [1] and its almost simultaneous explanation in terms of a hole-carrier-based pairing mechanism by Hirsch[2, 3] has triggered an avalanche of studies of its structural, magnetic and transport properties[4,5,6,7,8]. In spite of the significant advancement made on critical current density of HTS conductors the best transport $J_c$ of Ag/Bi2223 tapes is still limited by the grain connectivity. The study on the intergranular and intragranular critical current densities revealed that the former is smaller than the latter by at least one order of magnitude for the state-the-art Ag/Bi2223 tapes[9]. This characteristic feature of HTS conductors is clearly evident by the rapid drop of $J_c$ in low magnetic fields which is defined as weak link regime. Preliminary results on the newly discovered 40 K superconductor, $MgB_2$ reported by Larbalestier et al have indicated that $J_c$ of this material is not limited by grain boundary weak links as HTS[6]. The high critical current density of $10^5$ A/cm$^2$ at 11K and low field and $2\times10^4$ A/cm$^2$ at 20K and 1T have been reported on the bulk $MgB_2$ samples using magnetisation measurements[5,6]. As a further contribution to the field, we design a synthesis method to produce bulk $MgB_2$ samples with large and uniform grain size. The results of magnetisation measurements performed on these samples allow us to compare and evaluate the intergranular and intragranular critical currents in $MgB_2$.

## II. Experimental Details

Polycrystalline samples of $MgB_2$ were prepared by conventional solid state reaction. High purity Mg and B (amorphous) were mixed and finely grounded well, then pressed into pellets 10 mm in diameter with 1-2 mm thickness. Extra Mg was added in order to avoid Mg loss at high temperatures. These pellets were placed on an iron plate and covered with iron foil then put into a tube furnace. The samples were sintered at temperatures between 700 and 1000 $^oC$ for 1- 14 h. A high purity Ar gas flow was maintained throughout the sintering process. The sample purity was found to be very sensitive to oxygen contamination, any leaks in the furnace resulting in the oxidation of Mg to MgO. The purity of the Mg starting material was also a key factor influencing the quality of the resulting $MgB_2$

Phase purity was determined by XRD and grain sizes by SEM. Sintering time and temperature determine the phase content. The XRD pattern for a sample sintered for 4 h/800 $^o$ C is shown in Fig. 1, which shows mostly $MgB_2$ with only a small level of MgO contamination. The SEM results for the same sample are illustrated in Fig. 2. The grain size is of the order of 200 μm. It is likely that the large size grains were formed by a partially melting process. The sample appears to be dense, hard, and strong.

The transition temperature $T_c$ was determined by using AC susceptibility in an AC field of amplitude 1 Oe and frequency 117 Hz. The results, depicted in Fig.3, indicate a $T_c$ of 38.2 K with a very sharp transition, whose width of only 2K again confirms the high phase purity of the sample.

## III. Results and Discussion

**Magnetization and Flux Jimp**

The magnetization of a 3.5 x 2.5 x 1.3 $mm^3$ sample (cut from the initially prepared pellet) was measured over a temperature range of 5 to 35 K using a PPMS induction magnetometer in a time-varying magnetic field of sweep rate 50 Oe/s and amplitude 9 T. The magnetization results are depicted in Fig. 4. Immediately visible is a band of flux jumping, occurring over a narrow field range (H < 1T) at 5 and 10 K, that appears to truncate the maximum attainable magnetization. The Swartz and Bean sample size limitation on adiabatic flux jump stability[10] can be transformed into a magnetization limit of the form[11]:

$$M_{\max} = \frac{0.2}{3\pi}\sqrt{10^9 \frac{3}{\pi} \beta T^3 \Delta M_0(T)} \tag{1}$$

Where $\beta T^3$ represents the low temperature specific heat and $\Delta M_0(T)$ the magnetization temperature dependence in the form $\Delta M/(d\Delta M/dT)$. The lack of specific heat data prevents a direct prediction of $M_{max}$. However a back calculation using the experimental $M_{max}$ predicts a specific heat for $MgB_2$ at 5 K of $3x10^{-6}$ $J/cm^3$.

Evidence to be presented below indicates a sample, initially a bulk superconductor, that breaks down into a granular assembly beyond a certain critical value of field and temperature. Such a breakdown will also occur during a flux jump. The flux jumps depicted are partial, rather than complete. This suggests that a jump is terminated when the magnetization of the outer shell grains, and that of the residual core, drop below the critical value. Within the flux jump zone of Fig.4 (for T<10K and H<1T) we note the magnetization dropping to about one-fifth of its initial (virtual) value during the partial jump.

**Magnetic Critical Current Density**

A magnetic critical current density can be derived from the height of the magnetization loop $\Delta M$ using a suitable variant of the "semi-Bean" relationship $\Delta M = k J_c d$, where k is a constant and d the thickness that the sample presents to the applied field. $\Delta M$ itself versus H is plotted semi-logarithmically in Fig. 5. Here we note what are essentially a high-field and low-field regions of curves, the transition between which is complete at a fairly constant value of $\Delta M$. Since the low field family covers three orders of magnitude in $\Delta M$, the low-field sets are not visible in the usual linear type of M(H) loop (Fig.4).

The transition from low field step to high field step corresponds to a drop in $\Delta M$ by more than an order of magnitude. It is interesting to note that from the image of SEM of the sample the average grain size is less than one tenth of the sample size. It is evident that the drop in $\Delta M$ is attributable to the breakdown of grain matrix as a result of flux penetration to the grain boundaries which may contain impurities. In the low field region the current circulates mainly over the entire sample size, which is the intergranular current. The induced current circulates only in the individual grains in the high field region, which is the intragranular current [9]. The transition from low field step to high field step represents a transition from intergranular current regime to intragranular current regime.

Based on the full sample size at low fields and the grain size at high fields, the magnetic $J_c$ is calculated using the relationship for a plate in perpendicular field: $J_c = 20\Delta M/(a\text{-}a^2/3b)$. For low fields, the lateral dimensions of the sample are used: $a$=0.25cm, $b$=0.35cm, whereas for the high fields $a$=170 μm, $b$=260 μm. No calculation is applied to the intermediate-field transitional zone wherein the grain decoupling is taking place. The results are depicted in Fig. 6 wherein we note a continuous $J_c$ field dependence throughout the entire field range.

In summary, we presented evidence for decoupling of the grains in sintered $MgB_2$, both through partial flux jumping and a step in field-dependence of $J_c$. Similar step was also observed in the high-temperature superconductors. However a significant difference here is that the step occurs at much higher fields and it corresponds to a negligibly small drop of $J_c$ (Fig. 5). Therefore, grain connections in $MgB_2$ can sustain rather large $J_c$, in contrast to high-temperature superconductors. More significantly, the field dependence of inter-grain $J_c$ seems to be the same as that of intra-grain $J_c$ (Fig.6). The flux jump and step in $J_c(H)$ was observed thanks to large value of $J_c$ and large size of the grains. Our samples gave $J_c$ of $10^5$ A/cm$^2$ at 20 K and zero field, as obtained from magnetic hysteresis loops with sweep rate of field of 50 Oe/s. The value of $J_c$ would be much larger at lower temperature. However it was not possible to obtain it there from the hysteresis loop because of the flux jump.

**Acknowledgment:**

The authors would like to thank Australian Research Council for financial support.

Figure Captions:

Figure 1: XRD pattern for the sample showing that all the peaks can be indexed according to the published crystallography data of hexagonal $MgB_2$ except for the small peaks of MgO as impurity phase.

Figure 2: SEM images showing that the sample consists of large and uniform gains with size ranging from 0.15mm to 0.3mm.

Figure 3: AC susceptibility, measured with frequency and amplitude of ac field of 117Hz and 1 Oe, respectively.

Figure 4: Magnetic hysteresis loops, measured with the field sweep rate 50 Oe/s, at different temperatures

Figure 5: Width of the hysteresis loops, ΔM, as a function of field, with the sweep rate of the field 50Oe/s.

Figure 6: Critical current density vs. field, with corrections for different screening length of the currents in small and high fields.

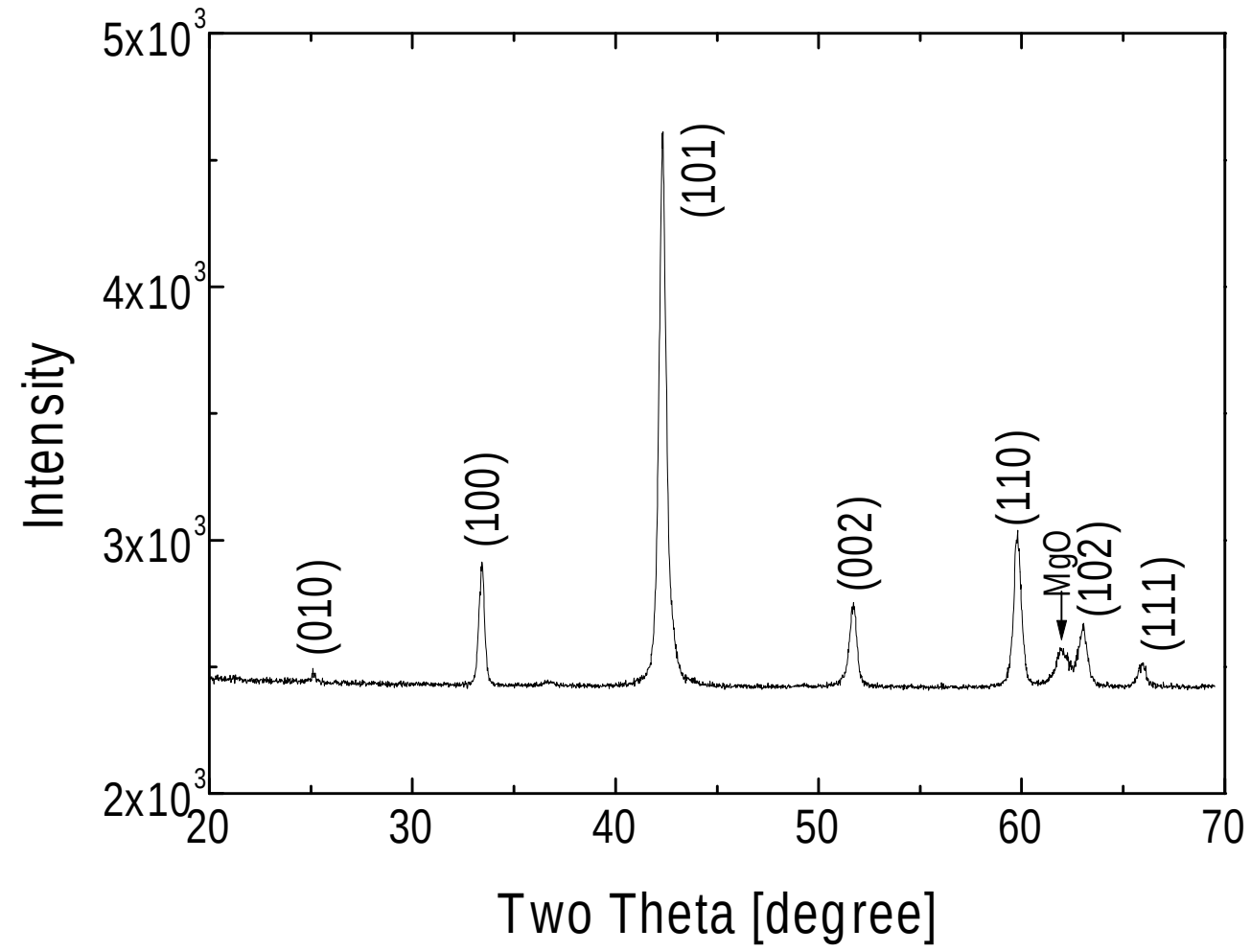

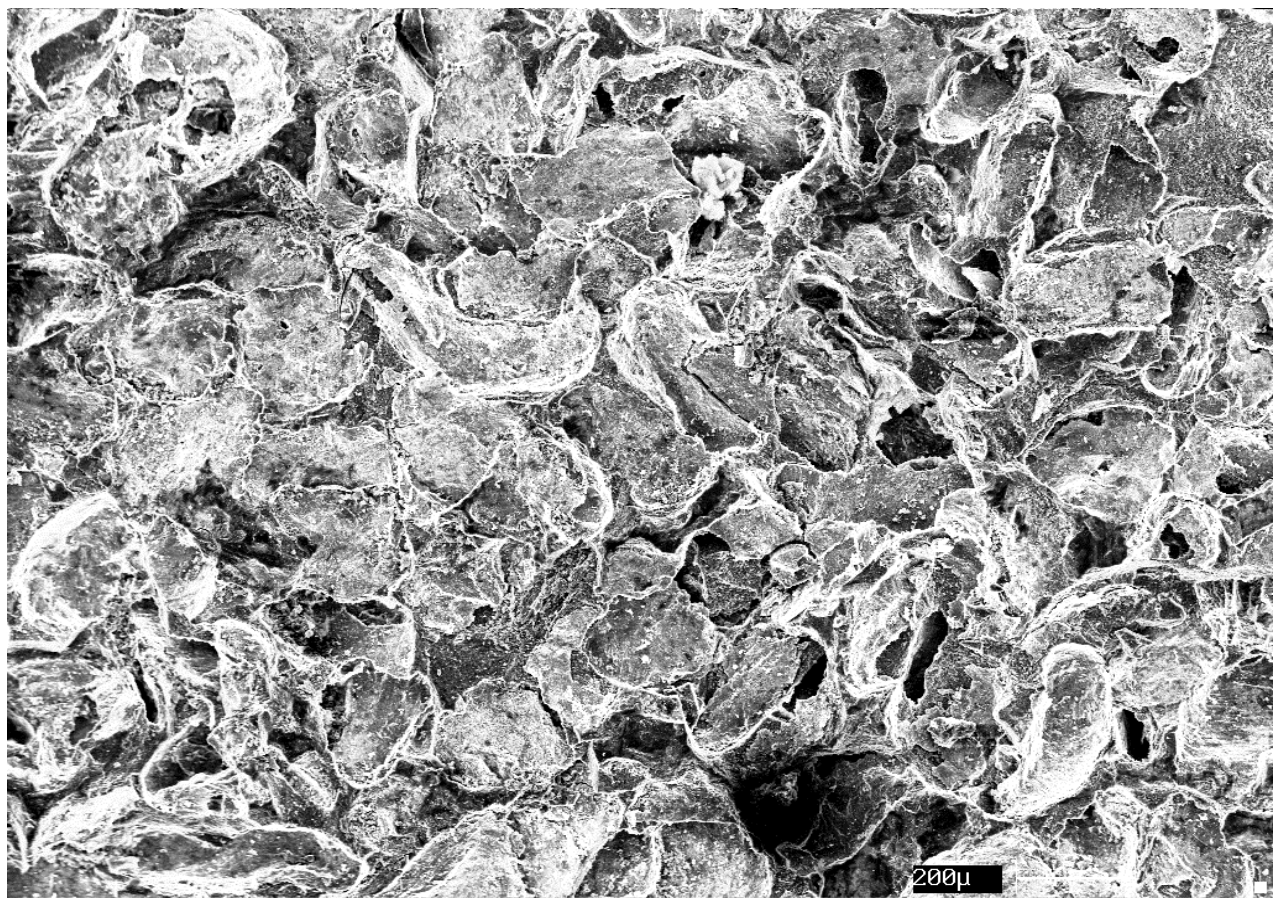

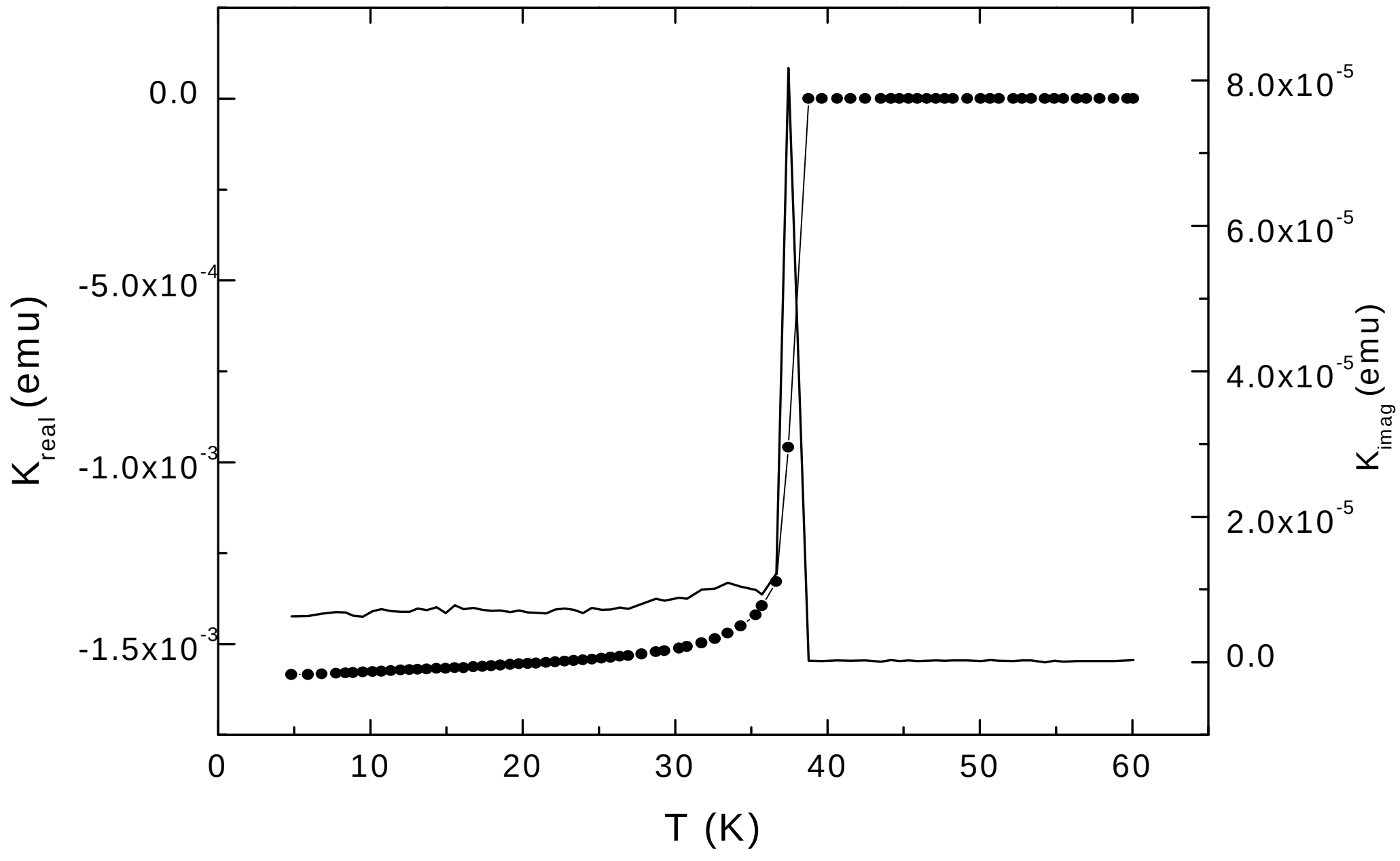

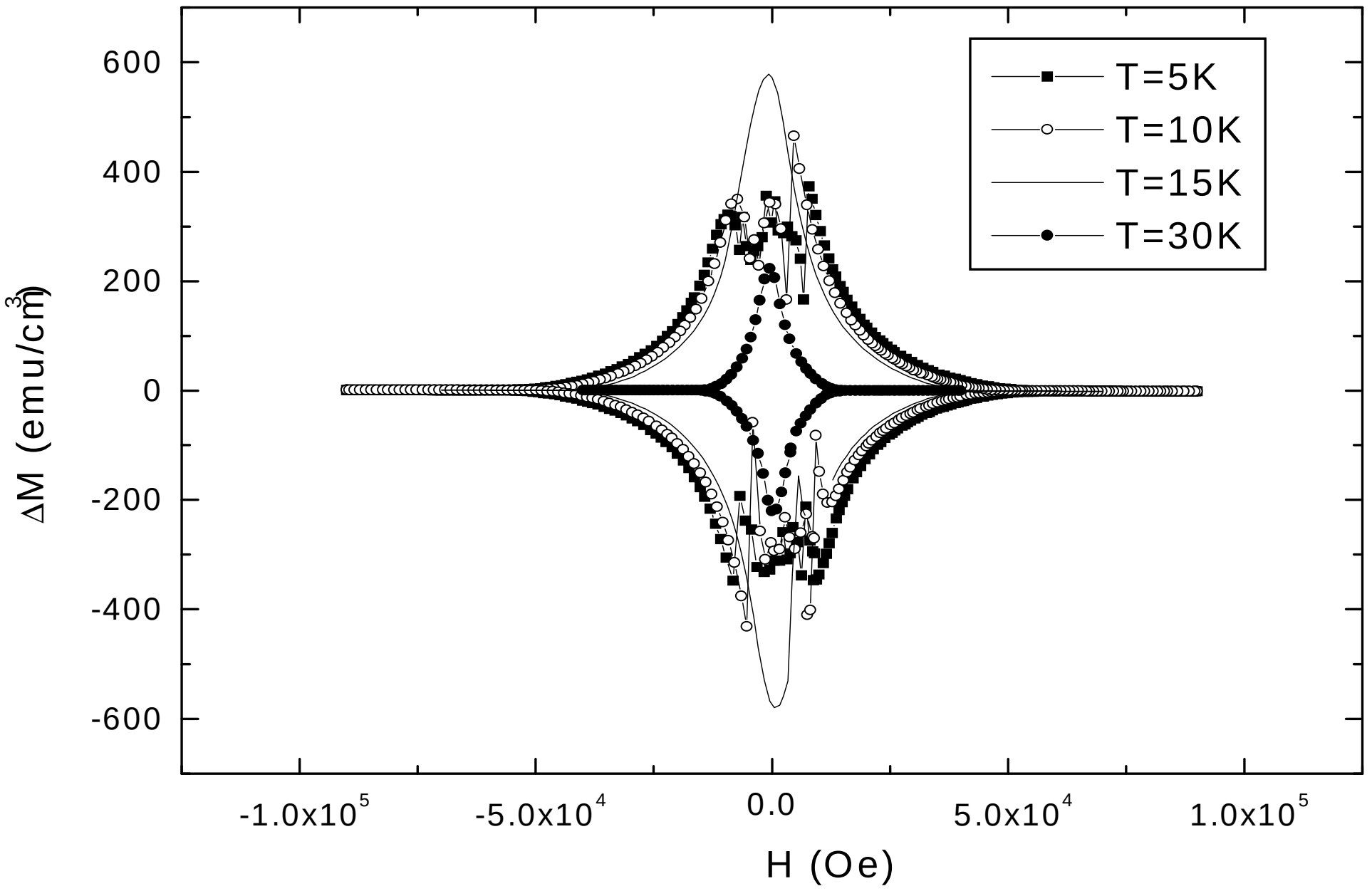

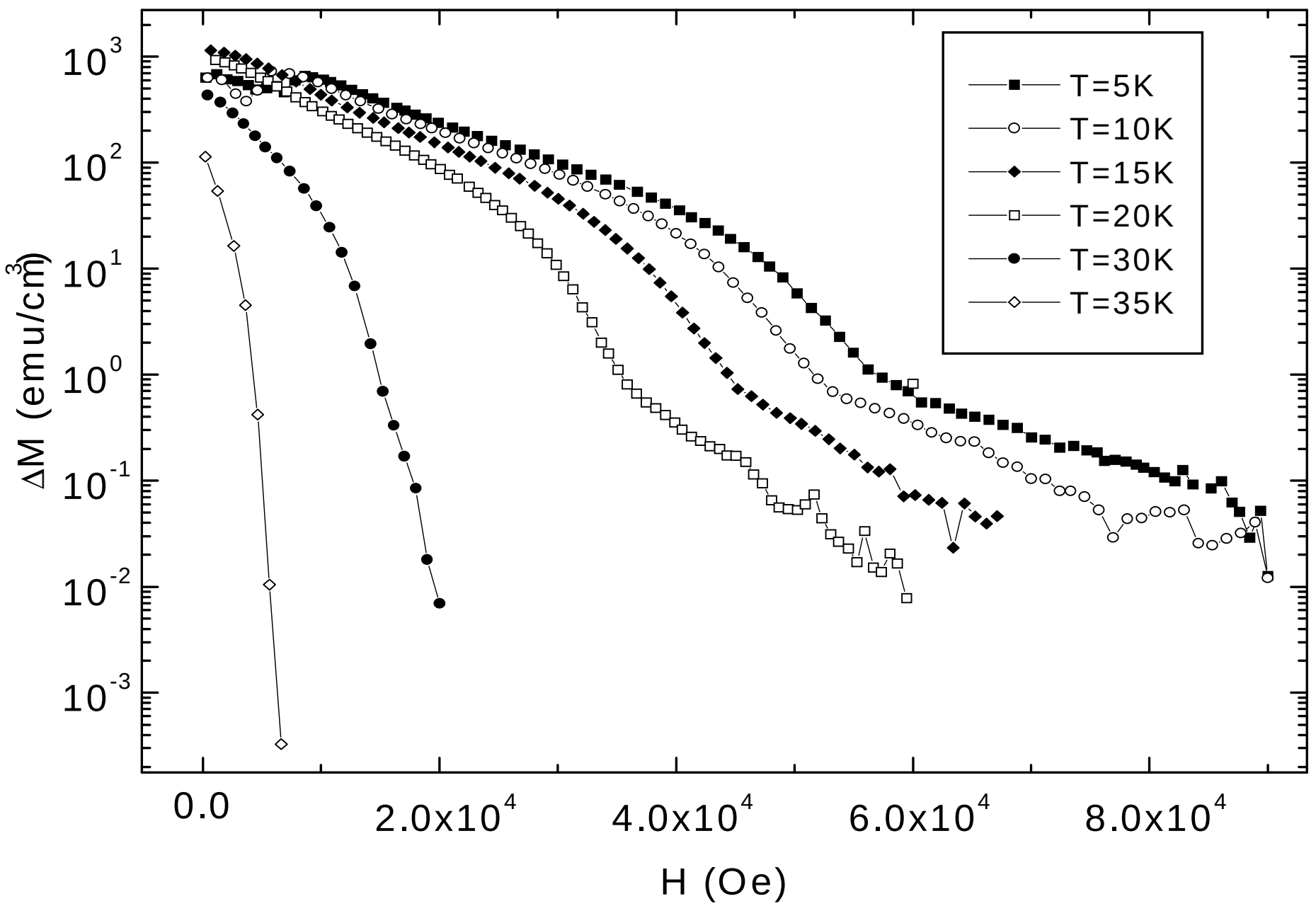

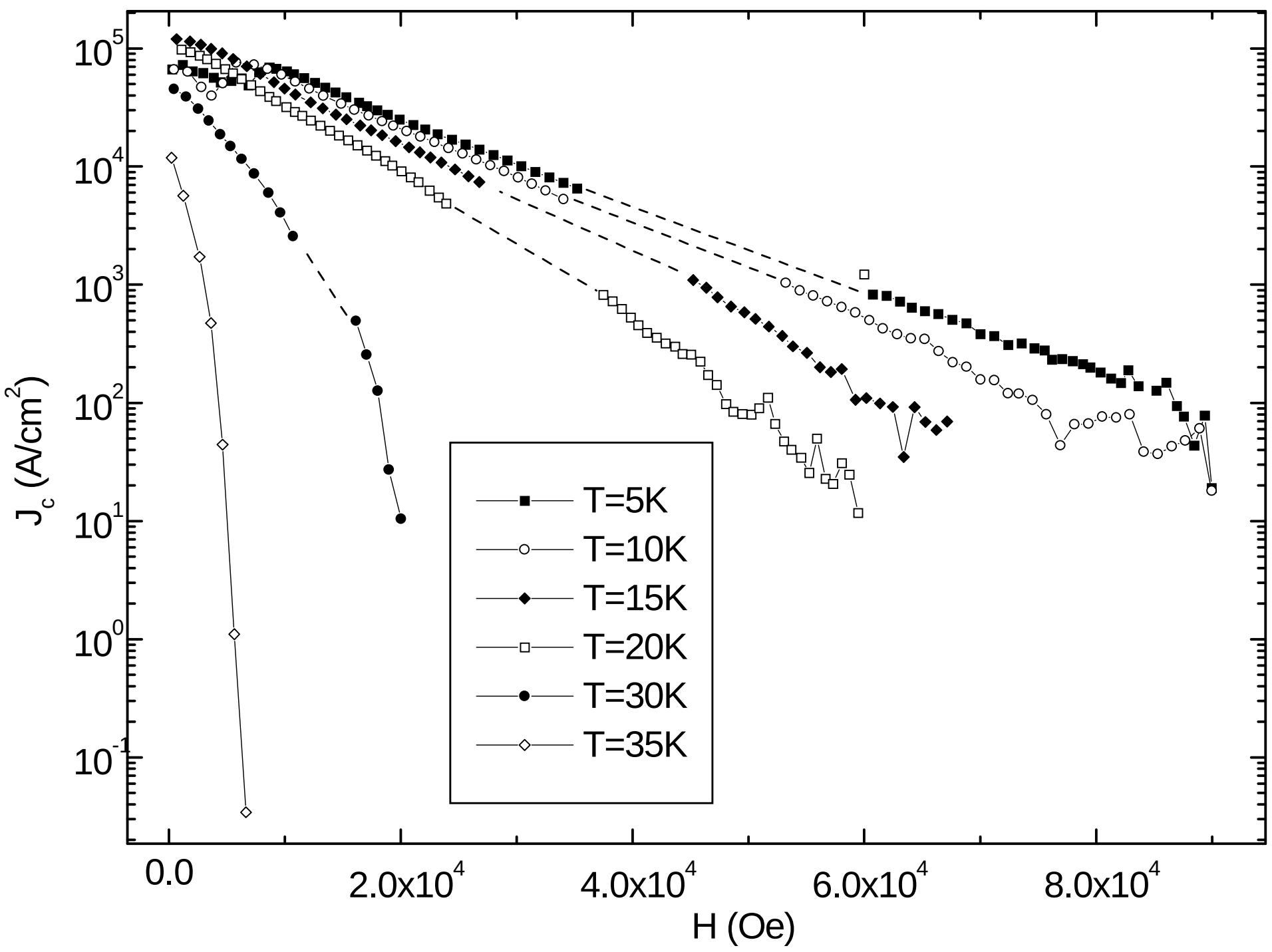